\RequirePackage{fix-cm}
\documentclass[smallextended]{svjour3}       
\smartqed  
\usepackage{graphicx}
\usepackage{mathptmx}      
\usepackage{latexsym}
\usepackage[version=3]{mhchem}
\usepackage[per=slash,output-decimal-marker={.},loctolang={DE:ngerman,UK:english},detect-weight=true,detect-family=true]{siunitx}
\usepackage[e]{esvect}
\usepackage{microtype}

\usepackage{color, soul}    

\newcommand\onemu{\mspace{1.5mu}}
\newcommand\of{\!\onemu}

\newcommand\deltarho{\mathord{\delta\!\rho}}
\newcommand\deltam{\mathord{\delta\! m}}

\newcommand\rhoofr{\rho\of(\vv{r})}

\DeclareSIUnit\cal{cal}
\DeclareSIUnit\kcal{\kilo\cal}
\journalname{Theoretical Chemistry Accounts}
\begin{document}

\title{Spin polarization in SCC-DFTB}


\author{Patrick Melix\and
        Augusto Faria Oliveira\and
        Robert R\"uger\and
        Thomas Heine
}


\institute{All Authors: \at
			  Wilhelm-Ostwald-Institut f\"ur Physikalische und Theoretische Chemie, Universit\"at Leipzig,\\
              Linn\'estr.~2, 04103~Leipzig, Germany\\
              Tel.: +49-351-9736403\\
              \email{patrick.melix@uni-leipzig.de}
              \and 
              	A.F.~Oliveira \and T.~Heine: \at 
                Department of Physics \& Earth Siences, 
                Jacobs University Bremen,\\ Campus Ring 1, 2859 Bremen, Germany
              \and
                Robert R\"uger: \at
                Scientific Computing \& Modelling NV,\\
                and Department of Theoretical Chemistry, Vrije Universiteit Amsterdam,\\ 
                De Boelelaan 1083, 1081 HV Amsterdam, The Netherlands
}

\date{Received: 4 May 2016 / Accepted: 23 August 2016}

\maketitle

\begin{abstract}
We evaluate the performance of spin-polarized DFTB within the SCC-DFTB (also known as DFTB2) model.
The method has been implemented in the ADF modeling suite.
We briefly review how spin polarization is incorporated into the DFTB2 method and validate the method in terms of structural parameters and energies using the \emph{GMTKN30} test set, from which we used 288 spin-polarized systems.
\end{abstract}

\section{Introduction}
\label{intro}
The density-functional-based tight-binding method (DFTB)~\cite{seifert1986} is an approximation of the Kohn-Sham density-functional theory (KS-DFT)~\cite{Hohenberg1964,Kohn1965} within the Linear Combination of Atomic Orbitals (LCAO) ansatz.
Since its introduction in the 1980s as a non-self-consistent approach~\cite{seifert1986}, two major extensions of DFTB have been developed, increasing its accuracy and widening the range of systems to which DFTB can be applied. 
The first -- and perhaps mostly used extension -- is the \emph{self-consistent charge correction} (SCC)~\cite{slako-mio}, which accounts for intramolecular charge transfer within the calculated systems due to the different chemical potentials and hardnesses of the atoms. 
Extending the method even further, the DFTB3~\cite{Gaus2011} describes, in addition to the SCC, changes in the chemical hardness of the atoms according to their electronic states.
Recently, the three DFTB variants have been referred to as DFTB1, DFTB2, and DFTB3, respectively~\cite{Elstner2014}, as they are derived from first-, second-, and third-order Taylor expansions of the Kohn-Sham energy functional. 
Although DFTB has originally been formulated for closed-shell systems, further developments have been made to include spin polarization effects in DFTB2~\cite{Frauenheim2000,Koehler2001,Koehler2005,Koehler2007}.

While the standalone DFTB+ program~\cite{DFTB+} is often considered the reference implementation of DFTB, the method has also been integrated into several program suites~\cite{demon,gaussian,atomistix,Koskinen2009,Rurali2003}.
The ADF modeling suite~\cite{SCM} also includes a DFTB implementation which supports DFTB1, DFTB2, and DFTB3 for periodic and finite closed-shell systems, and is closely coupled to its DFT code.
For many purposes in chemistry, having a common code basis for DFT and DFTB is advantageous, since DFTB results need to be validated and some properties might need better accuracy than DFTB calculations can provide.
Furthermore, the close integration between DFT and DFTB allows hybrid methods that selectively apply DFTB approximations in a DFT framework~\cite{Rueger2016}.
Finally, it allows the generation and validation of DFTB parameter sets within an common platform.
This integration has made the development of the \emph{QUASINANO} parameters~\cite{Wahiduzzaman2013,Oliveira2015} possible, which represents a significant step to resolve the limitation of DFTB to only a small part of the periodic table.
However, spin polarization has so far been missing in ADF-DFTB and a thorough benchmark highlighting the importance (or the lack of it) for light-weight molecules is not available in the literature.
In this paper, we first present the implementation of DFTB into the ADF-DFTB software. Then, we validate the method in terms of structural parameters and energies for an extensive set of spin-polarized molecules, i.e. for 288 systems found in the \emph{GMTKN30} reference set~\cite{Goerigk2010,Goerigk2011}.

The remainder of this article is organized as follows:
In section~\ref{method} we describe spin-polarized DFTB2 as implemented in ADF.
Next, we present reference data and the procedure used to validate our implementation, followed by a discussion of the results achieved.
Finally, we present our conclusions and final remarks. 
\section{Spin-Polarized DFTB}
\label{method}
In this section, the spin-polarized DFTB2 (a.k.a. SCC-DFTB) is described as implemented in ADF. 
For a general description of the DFTB method and its extensions, several review papers can be found in the literature~\cite{Elstner2014,Koskinen2009,Oliveira2009}.

The original spin-polarized DFTB method was presented in references~\cite{Frauenheim2000,Koehler2001}. In this original proposition, the spin polarization terms are calculated from single-atom contributions (one-center approximation). The implementation described here is based on the later work presented in references~\cite{Koehler2005,Koehler2007}, in which the spin-polarization terms have been extended into a two-center approximation.

The total energy in DFTB2 is given as~\cite{slako-mio,Frauenheim2002a}
\begin{equation}
   \begin{split}
      E^\mathrm{DFTB2}_\mathrm{tot}=& E_\mathrm{rep}+ \underbrace{\sum^{M}_{i}f_{i}\left\langle\psi_{i}\left|\hat{H}_0\of[\rho_0]\right|\psi_{i}\right\rangle}_{E_1}
      + \underbrace{\frac{1}{2}\sum^N_{A}\sum^N_{B} q_{A}q_{B} \gamma_{A,B}}_{E_\mathrm{2nd}=E_{\deltarho}}.
\end{split}
   \label{eq:dftb2_totalenergy}
\end{equation}
In equation~\eqref{eq:dftb2_totalenergy}, \(M\) denotes the number of orbitals in the system, \(N\) the number of atoms, \(f_i\) the occupation number of orbital \(i\), and
\(\hat{H}_0\) the reference Hamiltonian. The reference density \(\rho_0\), which defines \(\hat{H}_0\), is approximated as a sum of reference atomic electron densities as \(\rho_0=\sum^N_A \rho_0^A\) while the \(\rho_0^A\) contributions are calculated self-consistently for spherically symmetric spin-unpolarized neutral atoms.
The second order term \(E_\mathrm{2nd}\) contains the atomic Mulliken charges \(q\) and a function \(\gamma_{A,B}\)
which is an approximation for the charge transfer~\cite{Gaus2011}.
\(E_\mathrm{rep}\) is the repulsion energy term, which corresponds to the DFT \emph{double-counting} contributions plus the internuclear interactions. 
Usually, \(E_\mathrm{rep}\) is approximated as a sum of two-center potentials fitted to the difference between the full DFT energies and the sum of \(E_1\) and \(E_\mathrm{2nd}\) as polynomial or spline functions of the interatomic distances~\cite{Oliveira2009}.
However, there are alternative approaches in which \(E_\mathrm{rep}\) can be explicitly calculated~\cite{Oliveira2015,Mirtschink:2009tu,Bodrog:2012wx}.

In order to include collinear spin polarization (i.e. electron spins are parallel or anti-parallel to the direction of the external magnetic field)
in DFTB2, the total electron density is split into two electron spin densities~\cite{Koehler2005,Koehler2007}
\begin{equation}
      \rhoofr=\rho_\uparrow\of(\vv{r})+\rho_\downarrow\of(\vv{r})
      \label{eq:spindensity},
\end{equation}
resulting in a new quantity, namely the magnetization density 
\begin{equation}
      m\of(\vv{r})=\rho_\uparrow\of(\vv{r})-\rho_\downarrow\of(\vv{r})\label{eq:magnetizationdensity}.
\end{equation}
Similarly to the electron density, the magnetization density is described as the sum of a reference magnetization \(m_0\) and a fluctuation \(\deltam\),
with the difference that \(m_0\) is chosen to correspond to the spin-unpolarized reference density \(\rho_0\), which leads to \(m_0=0\).

Following the standard DFTB2 model~\cite{Elstner2014}, the total energy therefore becomes
\begin{equation}
     E\of[\rho,m]=
      	\left\lbrace	E_\mathrm{rep}\of[\rho_0,m_0]+E_1\of[\rho_0,m_0]+E_\mathrm{2nd}\of[\rho_0,\deltarho,m_0,\deltam]
      	\right\rbrace_{m_0=0},
	\label{eq:energy_dftbfinal}
\end{equation}
where the second-order energy \(E_\mathrm{2nd}\) is the only term depending on the density and magnetization fluctuations \(\deltarho\) and \(\deltam\). 
In addition, because the reference electron density \(\rho_0\) is unpolarized (i.e., $m_0=0$), \(\rho_0\), \(E_1\) and \(E_\mathrm{rep}\) are calculated exactly as in the DFTB1 model. Moreover, if the magnetization \(\deltam\) vanishes, equation \eqref{eq:energy_dftbfinal} becomes identical to the spin-unpolarized DFTB2 model~\cite{slako-mio}.

The second-order energy term in equation~\eqref{eq:energy_dftbfinal} can be further approximated as
\begin{equation}
   E_\mathrm{2nd}=E_{\deltarho}+E_{\deltam}=E_{\deltarho}+\frac{1}{2}\int\left.\frac{\delta^2 E_{\mathrm{XC}}}{\deltam^2}\right|_{\rho_0,m_0=0}\deltam^2\mathrm{d}^3r,
\end{equation}
in which \(E_{\deltam}\) describes the energy contribution due to the spin polarization. 
The magnetization density fluctuation is approximated  with a linear combination of atom-centered, spherically symmetric, and non-overlapping functions \(f\) as
\begin{equation}
   \deltam\of(\vv{r})=\sum_{A}^N p_{A}f_{A}\of(|\vv{r}-\vv{R}_{A}|),
\end{equation}
where \(N\) is the number of atoms and \(p_{A}\) is the difference between the spin up and spin down Mulliken populations of atom \(A\). Within this approximation, the second-order energy term can be rewritten as
\begin{equation}
   E_\mathrm{2nd}=E_{\deltarho}+\frac{1}{2}\sum^N_{A} p_{A}^2 W_{A}.
\end{equation}
The \(W_{A}\) variable is an atomic constant which can be calculated using the second derivative of the DFT energy of the free, spin-unpolarized atom. 
Using Janak's theorem~\cite{Janak1978} it can be formulated as
\begin{equation}
   W_{A}=\frac{1}{2}\left(\frac{\delta\epsilon_{\uparrow}}{\delta f_{\uparrow}}-\frac{\delta\epsilon_{\uparrow}}{\delta f_{\downarrow}}\right)
   \label{eq:magHub},
\end{equation}
with \(f\) being the occupation number and \(\epsilon\) the energy of the \emph{highest occupied atomic orbital}~(HOAO).

Applying the same reasoning to the Hamiltonian and the forces (relevant for geometry optimization) results in similar terms.
The Hamiltonian matrix elements are extended into
\begin{equation}
   \begin{split}
      {H}_{\uparrow/\downarrow,\mu \nu} =& {H}^0_{\mu \nu}+ \frac{1}{2}{S}_{\mu \nu}\sum^N_{C}\left(\gamma_{{A}\of(\mu),{C}}+\gamma_{{B}\of(\nu),{C}}
      \right)q_{C}\\
      &\pm\frac{1}{2}{S}_{\mu \nu}\left(W_{{A}\of(\mu)}\, p_{{A}\of(\mu)}+W_{{B}\of(\nu)}\, p_{{B}\of(\nu)}\right),
   \end{split}
\end{equation}
where \({A}\of(\mu)\) denotes the atom to which orbital \(\mu\) belongs. The forces translate into~\cite{KoehlerDoktor}
\begin{equation}
   \begin{split}
      F_{C}=&-\sum_{\sigma=\uparrow,\downarrow}\sum^\mathrm{occ.}_if_{i,\sigma}\sum_{\mu \nu}c^*_{\mu i \sigma}c_{\nu i \sigma}\left\{\frac{\delta {H}^0_{\mu \nu}}{\delta \vv{R}_{C}}+
      \frac{\delta {S}_{\mu \nu}}{\delta \vv{R}_{C}}\left(\frac{1}{2}\sum^N_{D}\left(\gamma_{{A}\of(\mu),{D}}
      +\gamma_{{B}\of(\nu),{D}}\right)q_{{D}}\right. \right.\\[2ex]
      &\left.\left.\pm\frac{1}{2}\left(W_{{A}\of(\mu)}\, p_{{A}\of(\mu)} +
      W_{{B}\of(\nu)}\, p_{{B}\of(\nu)}\right)-\epsilon_{i,\sigma}\right)\right\}
      -q_{C}\sum_{B}^Nq_{{B}}\frac{\partial\gamma_{{C},{B}}}{\partial\vv{R}_{C}}-\frac{\delta E_\mathrm{rep}}{\delta\vv{R}_{C}},
   \end{split}
\end{equation}
where \(\vv{R}_C\) are nuclear coordinates. 

In contrast to the \emph{orbitally-resolved} method shown in references~\cite{Koehler2005} and~\cite{Koehler2007}, the implementation presented in this paper corresponds to an \emph{atomically-resolved} approach. 
In the former, the \(\gamma\) and \(W\) parameters depend not only on the type of atoms involved, but also on the type of valence orbitals; hence, the contributions involving \(\gamma\) and \(W\) in the equations above would require the summation over the angular quantum number \(l\) of the valence orbitals, in addition to the summation over the atom indices. 

\section{Method Validation}
\label{validation}
We have used the \emph{GMTKN30} test set~\cite{Goerigk2010,Goerigk2011} to validate the spin-polarized DFTB2 method as implemented in ADF-DFTB. 
This test set consists of 30 subsets that include different types of reaction energies (e.g., ionization, isomerization, etc.), as well as structural parameters of different organic and inorganic species, including molecules, ions, and radicals.
The references are experimental or theoretical in origin, depending on the subset.
Thus, the \emph{GMTKN30} test set is suitable for testing both the accuracy and transferability of spin-polarized DFTB2.

However, most of the \emph{GMTKN30} subsets are completely composed of species in singlet states (closed-shell systems). 
Hence, we have only used seven of the subsets, since they include a significant amount of non-singlet calculations.
The subsets used are shown in table~\ref{table:subsets}; the number of reactions in each specified subset is given in the last column.

\begin{table}[h]
   \centering
   \caption{Subsets of the GMTKN30 test set~\cite{Goerigk2011} used in this work.}
   \label{table:subsets}
   \begin{tabular}{|c|p{0.65\linewidth}|c|}
      \hline
      \textbf{Subset} & \textbf{Description} & \textbf{Processes}  \\ \hline
       BH76 & barrier heights of hydrogen transfer, heavy atom transfer, nucleophilic substitution, uni molecular, and association reactions & 76 \\ \hline
       BH76RC & reaction energies of the BH76 set & 30 \\ \hline
       G21EA & adiabatic electron affinities & 25 \\ \hline
       G21IP & adiabatic ionization potentials & 36 \\ \hline
       MB08-165 & decomposition energies of artificial molecules & 165 \\ \hline
       RSE43 & radical stabilization energies &43 \\ \hline
       W4-08 & atomization energies of small molecules &99 \\ \hline
   \end{tabular}
\end{table}

It is important to note, that the goal of running these test sets is not to show good conformity of DFTB with the reference.
DFTB is an approximate method and one should therefore not expect perfect agreement with the reference.
The aim here is therefore to evaluate the results with and without spin polarization. 
Moreover, this is a good opportunity to compare results obtained with the \emph{3ob-3-1}~\cite{slako-3ob,slako-3ob2,slako-3ob3,slako-3ob4} and the recently published \emph{QUASINANO2015}~\cite{Oliveira2015} parameters.

The spin polarization parameters (table \ref{table:magHub}) have been calculated for the \emph{QUASINANO2015} parameter set using eq.~\eqref{eq:magHub}. 
The electronic eigenvalues have been calculated with the PBE exchange-correlation functional~\cite{pbe1996}, QZ4P basis sets, and scalar relativistic correction (ZORA)~\cite{zora1999}, as implemented in ADF~\cite{SCM}. 
Since the \emph{3ob-3-1} parameter set does not include spin-polarization parameters, the \(W\) values shown in table \ref{table:magHub} have been used.

\begin{table}[h]
\centering
\caption{Spin polarization parameters \(W\) in units of \(10^{-2}\)~hartree.}
\label{table:magHub}
\begin{tabular}{|cc|cc|cc|}\hline
\textbf{Element} & \textbf{\emph{W}} & \textbf{Element} & \textbf{\emph{W}} & \textbf{Element} & \textbf{\emph{W}} \\\hline
  H & -7.17 & O & -2.79 &  P & -1.49 \\\hline
  He & -8.66 & F & -2.99 &  S & -1.55 \\\hline
  Li & -1.98 & Ne & -3.17 &  Cl & -1.61 \\\hline
  Be & -2.30 & Na & -1.52 &  Ar & -1.66 \\\hline
  B & -1.96 & Mg & -1.66 &  K & -1.07 \\\hline
  C & -2.26 & Al & -1.40 &  Ca & -1.20 \\\hline
  N & -2.54 & Si & -1.44 &  Br & -1.38 \\ \hline
\end{tabular}
\end{table}

The subsets from table~\ref{table:subsets} underwent geometry optimization with the \emph{3ob-3-1} and \emph{QUASINANO2015} parameters using DFTB2 in ADF, in each case with and without spin polarization, resulting in four sets of results.
Orbitals were occupied according to a Fermi-Dirac distribution with a temperature of \(\SI{1}{\kelvin}\). We used a Broyden charge mixing~\cite{BroydenMixer} in the SCC cycle, with a mixing parameter of \(0.2\).
Calculations that failed due to convergence problems were automatically restarted using a higher temperature (steps of \(\SI{50}{\kelvin}\)) for orbital filling 
and a lower mixing parameter (steps of \(-0.01\)). The maximum number of retries was set to ten so that the maximal Fermi temperature was \(\SI{501}{\kelvin}\) and the minimal mixing parameter \(0.1\).

From the individual calculations, the energies of the appropriate reactions and the differences to the reference energies were calculated.
For failed calculations, the corresponding processes were not taken into account for comparison of the four test sets in terms of energy.
The main source of failed calculations were convergence problems and missing DFTB parameters (the \emph{3ob-3-1} parameter set contains less chemical elements than the \emph{QUASINANO2015} set and neither span all elements present in the \emph{GMTKN30} test subsets used in this work).

In addition to reaction energies, we have analyzed the mean absolute percent deviations~(MAPD) of calculated bond lengths and angles with respect to the \emph{GMTKN30} reference structures. 

\section{Results and Discussion}
\label{results}

In this section, we present and discuss a few representative results. Full test results can be found in the \emph{supporting information}.

Large deviations in the internal structures mainly appeared in the sets dealing with artificial molecules and transition states (BH76, BH76RC and W4-08, see table~\ref{table:subsets}).
All transition states in the BH sets transformed into the corresponding educts or products of the simulated reactions during the geometry-optimization.
Taking into account the approximate nature of the DFTB method, this is not surprising and therefore the bad conformity for these test sets should not be overinterpreted.

\begin{figure*}[htp]
   \centering
   \includegraphics[width=0.85\textwidth,keepaspectratio]{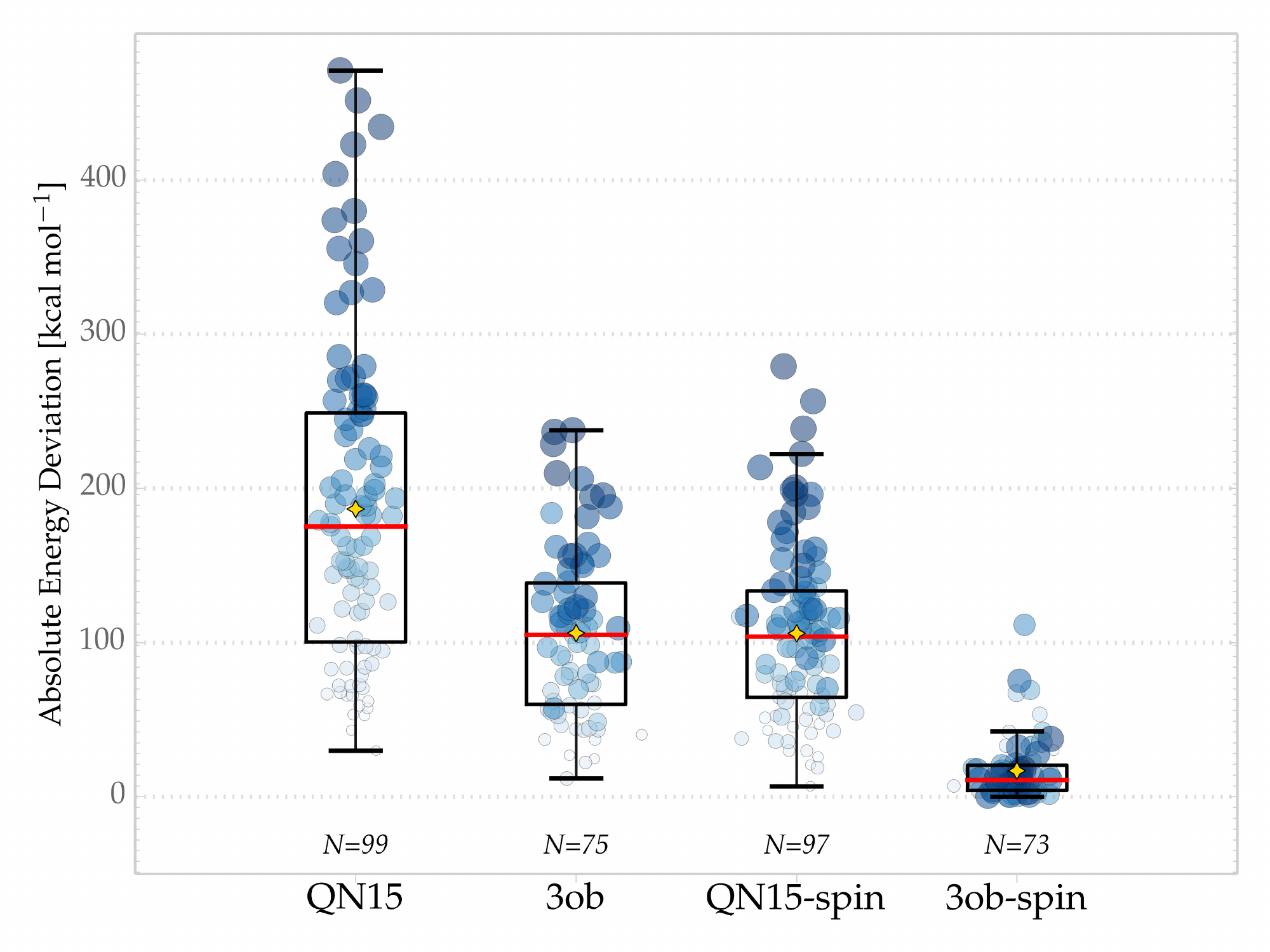}
   \caption{Boxplot of the absolute atomization energy deviations of the W4-08 subset using \emph{3ob-3-1} (\emph{3ob}) and \emph{QUASINANO2015} (\emph{QN15}) parameters without and with spin polarization.
Yellow stars denote mean values, red lines the median of the values, the whiskers span the last points within \(1.5\cdot\)IQR below the first quartile and above the third quartile, and the boxes span the the range between the first and third quartiles.
The number of data points is given as \(N\).}
   \label{fig:W4-08energy}
\end{figure*}

\begin{figure*}[htp]
   \centering
   \includegraphics[width=0.85\textwidth,keepaspectratio]{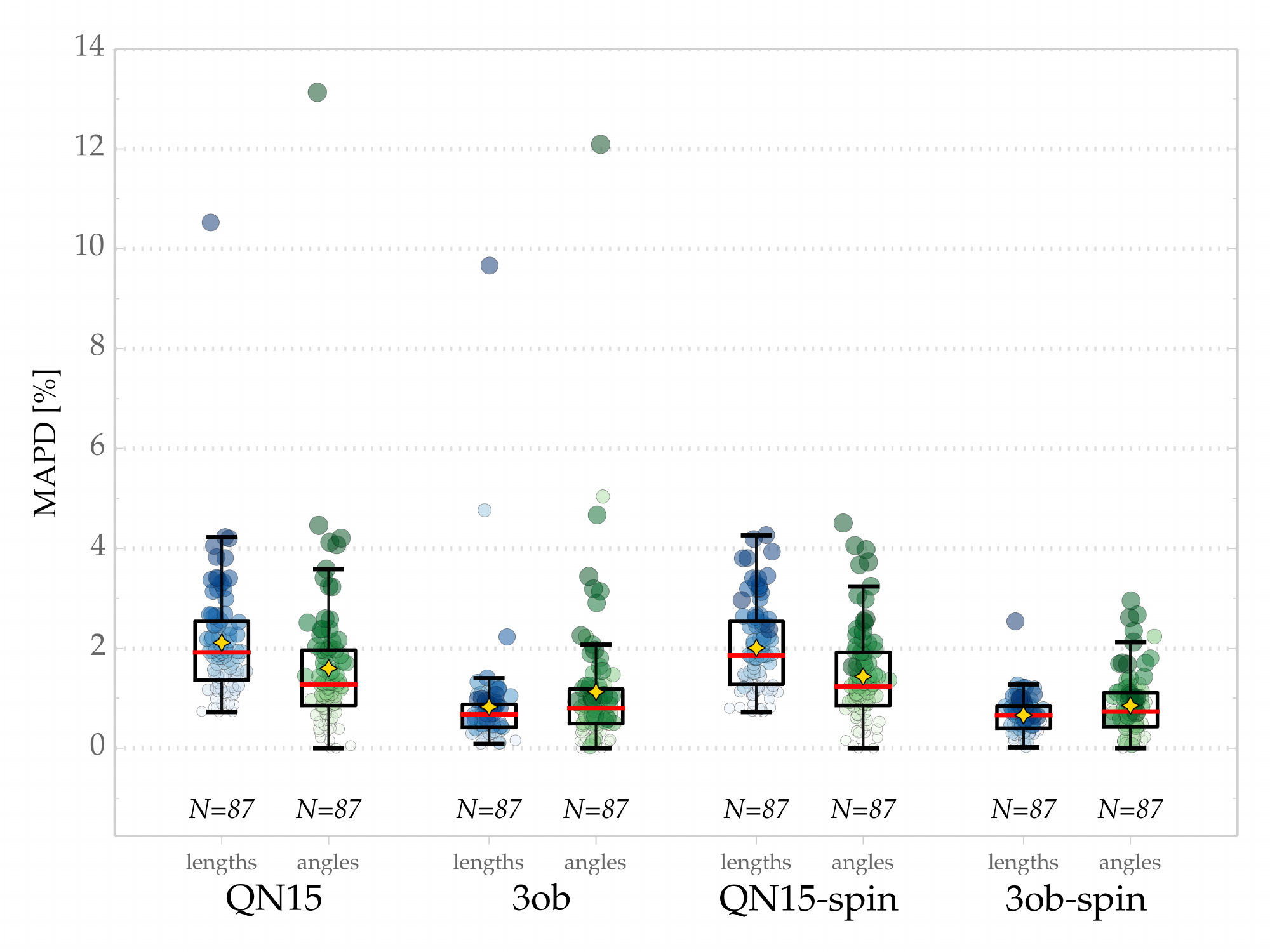}
   \caption{Boxplot of the mean-absolute percentage deviation (MAPD) of bond lengths (blue) and angles (green) in the RSE43 subset.
   Yellow stars denote mean values, red lines the median of the values, the whiskers span the last points within \(1.5\cdot\)IQR below the first quartile and above the third quartile, and the boxes span the the range between the first and third quartiles.
The number of data points is given as \(N\).}
   \label{fig:RSE43geom}
\end{figure*}

As a representation of the energetic results obtained, the absolute deviations of atomization energies are given in figure~\ref{fig:W4-08energy}. As expected, in the case of open-shell systems (which includes most atoms), the energies are heavily influenced by the spin polarization terms.

Also as expected, most geometries do not differ much with or without spin polarization. The only exception arises in the RSE43 subset, dealing with radicals. 
As can be seen in figure~\ref{fig:RSE43geom}, one of those radicals yields very different geometries when turning on the spin polarization, namely the \ce{Cl3C-C^.H2} radical. 

When calculated with spin polarization, the \ce{Cl3C-C^.H2} structure remains stable; in the spin-unpolarized calculations, however, one of the Cl atoms relocates to the other carbon and the structure rearranges into \ce{Cl2C^.-CClH2}. 
Further calculations with spin polarization and the \emph{3ob-1-1} parameters have shown that the energy of the rearranged structure (\ce{Cl2C^.-CClH2}) is in fact lower than the energy of the \ce{Cl3C-C^.H2} radical by ca.~\(\SI{14}{\kcal\per\mol}\), which is not a very large difference within the accuracy of DFTB. 
Therefore, it is likely that the \ce{Cl3C-C^.H2} structure corresponds to a relatively shallow local minimum in the potential energy surface; in this case, a small destabilization of the electronic structure (which could easily happen by neglecting the spin polarization contribution) would be enough to either eliminate the local minimum or to make it too shallow to be detected by the geometry optimization algorithm, thus explaining the structural rearrangement.  
Nevertheless, finding a definitive answer still requires further investigation. 


All obtained results are available as raw data tables and plots for further investigation and comparison in the supporting information.
\section{Conclusion}
The ADF-DFTB implementation has been successfully updated to include spin-polarization terms necessary for the description of open-shell chemical species.
Using the \emph{GMTKN30} test set, a large number of molecules and atoms were successfully calculated, thus validating the usefulness of the new features.

As seen in the results, reaction energies are strongly influenced by spin-polarization terms when the systems contain unpaired electrons. The importance of spin polarization in DFTB is further evidenced by the structural deviations in the \ce{Cl3C-C^.H2} radical. Although the \ce{Cl3C-C^.H2} does not represent the majority of the results obtained in this work, it clearly demonstrates that spin polarization can deeply affect molecular geometries in certain cases and, hence, should not be neglected. 

The next steps in development of the code should now be to implement periodic boundary conditions and orbital dependency. 
These steps are necessary to finally be able to achieve the goal of creating effective DFTB parameter sets for the whole periodic table and thus remove one of the important limitations of DFTB at the moment.

\begin{acknowledgements}
We thank SCM for the ongoing and fruitful cooperation and the Center for Information Services and High Performance Computing (ZIH) at TU
Dresden for allocation of computer time. We also thank the European Union’s Seventh Framework Programme (FP7-PEOPLE-2012-ITN) under project PROPAGATE (GA 316897) and the Deutsche Forschungsgemeinschaft (DFG) for their financial support. 
\end{acknowledgements}




\end{document}